\documentclass[a4paper,11pt]{article}

\usepackage{amsmath,amssymb,graphicx,bigints,multirow,siunitx,booktabs,float}
\usepackage[usenames,dvipsnames]{xcolor}
\usepackage[margin=1.9cm]{geometry}
\usepackage[colorlinks = true,allcolors=blue]{hyperref}
\usepackage[numbers,sort&compress]{natbib}
\usepackage{sectsty}
\subsectionfont{\normalfont\bf\normalsize\underline}
\subsectionfont{\normalfont\bf\normalsize}
\newcommand{\revision}[1]{#1}
\usepackage[size=footnotesize]{caption}

\usepackage{tgpagella,eulervm}

\title{\revision{Modelling and analysis of laser flash experiments}\\ using the Cattaneo heat equation}
\author{Elliot J. Carr\footnote{Corresponding Author (\href{elliot.carr@qut.edu.au}{elliot.carr@qut.edu.au})}\\ \small School of Mathematical Sciences, Queensland University of Technology (QUT), Brisbane, Australia}

\date{}

\begin{document}
\maketitle

\begin{abstract}
\noindent Thermal diffusivity of solid materials is commonly measured using laser flash analysis. This technique involves applying a heat pulse to the front surface of a small sample of the material and calculating the thermal diffusivity from the resulting increase in temperature on the back surface. Current formulas for the thermal diffusivity are based on the assumption that heat is transported within the sample according to the standard heat equation. While this assumption is valid in most practical cases, it admits the non-physical property of infinite propagation speed, that is, the heat pulse applied at the front surface is instantaneously perceived at the back surface. This paper carries out a mathematical analysis to determine the effect of replacing the standard heat equation in laser flash analysis by the Cattaneo heat equation, which exhibits finite propagation speed through the inclusion of a relaxation time in the Fourier law. The main results of the paper include (i) analytical insights into the spatiotemporal behaviour of temperature within the sample and (ii) analytical formulas for determining the thermal diffusivity and relaxation time of the sample. Numerical experiments exploring and verifying the analytical results are presented with supporting MATLAB code made publicly available.
\end{abstract}

\section{Introduction}
\begin{figure}[t]
\centering
\fbox{\includegraphics[width=0.85\textwidth]{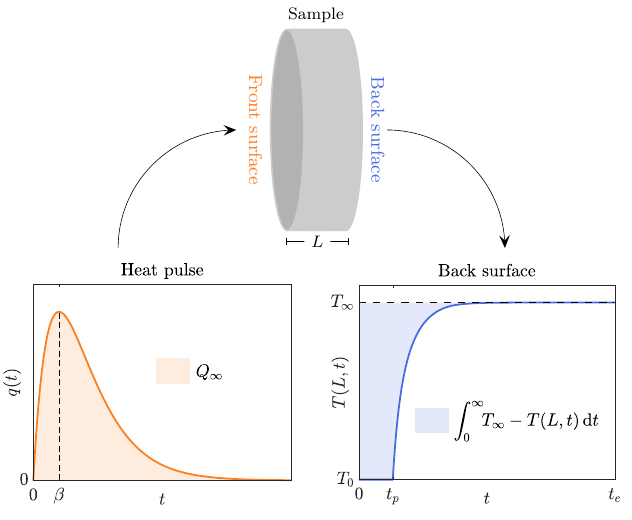}}
\caption{\textbf{Laser flash method with the Cattaneo heat equation.} A heat pulse is applied to the front surface of a thin sample of thickness $L$ with parallel front and back surfaces with $Q_{\infty} = \int_{0}^{\infty}q(t)\,\text{d}t$ denoting the total amount of heat applied. This causes the temperature on the back-surface to increase from an initial value of $T_{0}$ to a steady-state value of $T_{\infty}$. Due to the finite propagation speed of the Cattaneo heat equation, the back-surface temperature remains at the initial temperature of $T_{0}$ until $t = t_{p}$ and then increases.}
\label{fig:1}
\end{figure}

Laser flash analysis \cite{parker_1961} is the most common technique for measuring the thermal diffusivity of solid materials \cite{vozar_2003,czel_2013,blumm_2002}. This method involves subjecting the front surface of a small thermally-insulated sample of the material (usually thin and disc-shaped with parallel front and back surfaces) to a heat pulse and recording the temperature increase on the back surface (Figure \ref{fig:1}). Developing a heat transfer model to describe the temperature within the sample over time, allows formulas for the thermal diffusivity to be derived, expressed in terms of the time-dependent back-surface temperature history. Currently, such formulas \cite{carr_2019a,carr_2019b,carr_2023a,baba_2009} are derived based on the assumption that heat is transported within the sample according to the standard heat equation
\begin{gather}
\label{eq:heat_equation}
\frac{\partial T}{\partial t} = \alpha\frac{\partial^{2} T}{\partial x^{2}},
\end{gather}
where $\alpha$ is the thermal diffusivity and $T(x,t)$ is the temperature at position $x$ and time $t$. A drawback of this assumption is that the standard heat equation admits the non-physical property that heat \revision{is} propagated at infinite speed \cite{capriz_2021,madhukar_2019}. In the context of the laser flash method, this means that the heat pulse applied at the front surface is instantly perceived at the back surface, resulting in an immediate increase in the temperature at the back surface. While this assumption is valid for most metals it is invalid in some notable cases such as biological and polymeric materials \cite{reverberi_2008,mitra_2005}. The best known approach for overcoming this non-physical property is to replace the standard heat equation with the Cattaneo heat equation \cite{madhukar_2019,christov_2005}:
\begin{gather}
\label{eq:cattaneo_equation}
\frac{\partial T}{\partial t} + \tau\frac{\partial^{2} T}{\partial t^{2}} = \alpha\frac{\partial^{2} T}{\partial x^{2}},
\end{gather}
which includes an additional parameter, $\tau$, known as the relaxation time \cite{dangui-mbani_2017}. The Cattaneo equation exhibits a finite propagation speed of $s_{p}:=\sqrt{\alpha/\tau}$ \cite{barna_2010,ali_2005}, which means that the heat pulse is perceived at the back surface when $t = t_{p} := L\sqrt{\tau/\alpha}$, where $L$ is the thickness of the sample (Figure \ref{fig:1}). Note that the standard heat equation (\ref{eq:heat_equation}) is recovered from the Cattaneo equation (\ref{eq:cattaneo_equation}) when $\tau\rightarrow 0$ yielding an infinite propagation speed ($s_{p}\rightarrow\infty$) and an instant increase in the temperature at the back surface ($t_{p}\rightarrow 0$). The Cattaneo equation \revision{can be} derived by replacing the standard Fourier law for the heat flux, $\revision{\phi}(x,t) = -k\frac{\partial T}{\partial x}(x,t)$, where $k$ is the thermal conductivity, with the modified Fourier law \cite{angelani_2020,barletta_1997,pietrzak_2024}
\begin{gather}
\label{eq:modified_fourier}
\tau\frac{\partial \revision{\phi}}{\partial t} + \revision{\phi} = -k\frac{\partial T}{\partial x},\qquad \revision{\phi}(x,0) = 0\revision{.}
\end{gather}
\revision{In a similar manner to the standard heat equation, the Cattaneo equation can also be derived in the framework of continuum irreversible thermodynamics \cite{procopio_2024} as the continuum limit of a particular stochastic random walk process. As pointed out by many authors, the modified Fourier law (\ref{eq:modified_fourier})} can also be viewed as a first-order approximation to the delayed Fourier law 
$\revision{\phi}(x,t+\tau) = -k\frac{\partial T}{\partial x}(x,t)$ \cite{ozisik_1994,lopez_2014,blauth_2023}.  Solving (\ref{eq:modified_fourier}) yields the Cattaneo-form of the heat flux, which is non-local and involves an integral of the temperature gradient history:
\begin{gather*}
\revision{\phi}(x,t) = -\bigintssss_{\,0}^{t} \frac{k}{\tau}\exp\left(\frac{u-t}{\tau}\right)\frac{\partial T}{\partial x}(x,u)\,\text{d}u.
\end{gather*}

\revision{Non-Fourier heat conduction remains an ongoing and active research area. Recent work has focussed on both the dual-phase-lag model \cite{tzou_1995} and the Cattaneo Fourier law (or single-phase-lag model), with the former incorporating time delays into the temperature gradient as well as the heat flux. This recent work has included studies on thermo-visco-elastic interactions in a rotating micro-scale beam exposed to a laser pulse heat source \cite{abouelregal_2022}; magneto-thermo-elastic interactions in porous material with a spherical cavity exposed to laser pulse heating \cite{alsaeed_2024,abouelregal_2025}; and heat transfer in bone tissue subjected to a laser heat source \cite{katirachi_2023}. In each of these papers, the governing equations are solved using the Laplace transform method, with the transformed solution inverted back to the time domain numerically. Extensive reviews on non-Fourier heat conduction have also been published recently, focussing on bioheat transfer \cite{azhdari_2025} and functionally graded materials~\cite{amiri-delouei_2025}.}

In this paper, we carry out a mathematical analysis of the laser flash method applied to the Cattaneo heat equation (\ref{eq:cattaneo_equation}) \revision{for the idealised configuration of a homogeneous, isotropic, thermally insulated slab with parallel front and rear surfaces and uniform application of heat at the front surface (Figure~\ref{fig:1}). While we acknowledge that the assumption of perfect thermal insulation has limited practical use, the main aim of the work is to improve the fundamental understanding of the Cattaneo equation and uncover the role played by the relaxation time in thermal diffusivity calculations}. Key contributions include analytical insights into the spatiotemporal behaviour of the sample temperature (after application of the heat pulse) (section \ref{sec:average}--\ref{sec:exact_solution}) and analytical formulas for calculating the thermal diffusivity and relaxation time from the back-surface temperature history (sections \ref{sec:thermal_diffusivity}--\ref{sec:relaxation_time}). Numerical experiments supporting and verifying the analysis are presented in section~\ref{sec:implementation_results}. \revision{Modification of the analysis to incorporate heat losses from the front and rear surfaces is then discussed in section \ref{sec:heat_loss}} and the paper summarised and concluded in section~\ref{sec:conclusions}.

\section{Laser flash analysis}
\subsection{Cattaneo heat transfer model}
\label{sec:model}
In this paper, we assume the sample temperature satisfies the heat transfer model presented in \cite{carr_2019b}, modified appropriately to incorporate the Cattaneo heat equation:
\begin{gather}
\label{eq:model_phi_pde}
\frac{\partial T}{\partial t} + \tau\frac{\partial^{2} T}{\partial t^{2}} = \alpha\frac{\partial^{2} T}{\partial x^{2}},\\
\label{eq:model_phi_ics}
T(x,0) = T_{0},\quad \frac{\partial T}{\partial t}(x,0) = 0,\\ 
\label{eq:model_phi_bcs}
\revision{\phi}(0,t) = q(t),\quad \revision{\phi}(L,t) = 0.
\end{gather}
Here $T_{0}$ is the initial (uniform) temperature of the sample and $q(t)$ denotes the heat flux applied by way of the heat pulse at the front surface at time $t$, which we assume satisfies $q(0)=0$, $\lim_{t\rightarrow\infty}q(t)=0$ and $\lim_{t\rightarrow\infty}q'(t)=0$. Note that two initial conditions (\ref{eq:model_phi_ics}) are required due to the presence of the second-order time derivative in the Cattaneo equation (\ref{eq:model_phi_pde}). 

The modified Fourier law (\ref{eq:modified_fourier}) allows the boundary conditions (\ref{eq:model_phi_bcs}) to be expressed in terms of the standard Fourier flux, giving the following equivalent alternative model:
\begin{gather}
\label{eq:model_pde}
\frac{\partial T}{\partial t} + \tau\frac{\partial^{2} T}{\partial t^{2}} = \alpha\frac{\partial^{2} T}{\partial x^{2}},\\
\label{eq:model_ics}
T(x,0) = T_{0},\quad \frac{\partial T}{\partial t}(x,0) = 0,\\ 
\label{eq:model_bcs}
-k\frac{\partial T}{\partial x}(0,t) = \widetilde{q}(t),\quad \frac{\partial T}{\partial x}(L,t) = 0,
\end{gather}
where 
\begin{gather}
\label{eq:qtilde}
\widetilde{q}(t) := \tau q'(t) + q(t).
\end{gather}

\subsection{Average temperature}
\label{sec:average}
We now derive an analytical expression for the average temperature within the sample over time:
\begin{gather}
\label{eq:Tavg_def}
\overline{T}(t) = \frac{1}{L}\int_{0}^{L} T(x,t)\,\text{d}x.
\end{gather}
Integrating the Cattaneo equation (\ref{eq:model_pde}) from $x=0$ to $x=L$ and then dividing by $L$ yields
\begin{gather*}
\frac{\text{d}\overline{T}}{\text{d}t} + \tau\frac{\text{d}^{2}\overline{T}}{\text{d}t^{2}} = \frac{\alpha}{L}\left[\frac{\partial T}{\partial x}(L,t) - \frac{\partial T}{\partial x}(0,t)\right].
\end{gather*}
Using the boundary conditions (\ref{eq:model_bcs}) and applying the averaging operator to the initial conditions (\ref{eq:model_ics}), we see that $\overline{T}(t)$ satisfies the initial value problem:
\begin{gather*}
\frac{\text{d}\overline{T}}{\text{d}t} + \tau\frac{\text{d}^{2}\overline{T}}{\text{d}t^{2}} = \frac{\widetilde{q}(t)}{\rho c L},\\
\overline{T}(0) = T_{0},\quad \frac{\text{d}\overline{T}}{\text{d}t}(0) = 0.
\end{gather*}
This initial value problem can be solved exactly using standard techniques (e.g. variation of parameters) yielding the following solution:
\begin{gather*}
\overline{T}(t) = T_{0}+\frac{1}{\rho c L}\left[\int_{0}^{t}\widetilde{q}(u)\,\text{d}u - \int_{0}^{t}\exp\left(\frac{u-t}{\tau}\right)\widetilde{q}(u)\,\text{d}u\right]\!.
\end{gather*}
Using (\ref{eq:qtilde}) and recalling that $q(0) = 0$, it is straightforward to show that
\begin{gather*}
\int_{0}^{t}\widetilde{q}(u)\,\text{d}u = \tau q(t) + \int_{0}^{t}q(u)\,\text{d}u \qquad\text{and}\qquad
\int_{0}^{t}\exp\left(\frac{u-t}{\tau}\right)\widetilde{q}(u)\,\text{d}u = \tau q(t),
\end{gather*}
and hence the average temperature is given by
\begin{gather}
\label{eq:Tavg}
\overline{T}(t) = T_{0}+\frac{Q(t)}{\rho c L},
\end{gather}
where $Q(t) := \int_{0}^{t}q(u)\,\text{d}u$. Note that $\overline{T}(t)$ is independent of the relaxation time, $\tau$, and is equivalent to the expression obtained for the standard heat equation \cite{carr_2019b}. The form of $\overline{T}(t)$ (\ref{eq:Tavg}) also confirms conservation of heat for the Cattaneo heat transfer model (\ref{eq:model_pde})--(\ref{eq:model_bcs}) with the change in the amount of heat in the sample balanced by the amount entering at the front surface (i.e. $\smash{\rho c \int_{0}^{L} T(x,t) - T_{0}\,\text{d}x = Q(t)}$ can be obtained by combining (\ref{eq:Tavg_def}) and (\ref{eq:Tavg})).

\subsection{Steady-state temperature}
We now derive an analytical expression for the steady-state temperature field 
\begin{gather*}
T_{\infty} = \lim_{t\rightarrow\infty} T(x,t).
\end{gather*}
Using (\ref{eq:qtilde}) and recalling that $\lim_{t\rightarrow\infty}q(t)=0$ and $\lim_{t\rightarrow\infty} q'(t)=0$, we see that $T_{\infty}$ satisfies the boundary value problem:
\begin{gather*}
0 = \alpha\frac{\text{d}^{2} T_{\infty}}{\text{d} x^{2}},\\
\frac{\text{d} T_{\infty}}{\text{d} x}(0) = 0,\quad \frac{\text{d} T_{\infty}}{\text{d} x}(L) = 0.
\end{gather*}
This problem identifies $T_{\infty}$ as a constant, which can be identified by taking the long time limit of the average temperature, $\overline{T}(t)$ (\ref{eq:Tavg}), i.e. $T_{\infty} = \lim_{t\rightarrow\infty} \overline{T}(t)$, giving:
\begin{gather}
\label{eq:Tinf}
T_{\infty} = T_{0}+\frac{Q_{\infty}}{\rho c L}, 
\end{gather}
where $Q_{\infty} := \lim_{t\rightarrow\infty}Q(t) = \int_{0}^{\infty} q(u)\,\text{d}u$. As with $\overline{T}(t)$, we note that $T_{\infty}$ is independent of the relaxation time, $\tau$, and is equivalent to the expression obtained for the standard heat equation \cite{carr_2019b}.

\subsection{Exact solution for temperature in space and time}
\label{sec:exact_solution}
We now develop an analytical solution of the Cattaneo heat transfer model (\ref{eq:model_pde})--(\ref{eq:model_bcs}). Taking the Laplace transform of equations (\ref{eq:model_pde})--(\ref{eq:model_bcs}) yields the following boundary value problem:
\begin{gather*}
(1+\tau s)(sT_{\mathcal{L}}-T_{0}) = \alpha\frac{\text{d}^{2}T_{\mathcal{L}}}{\text{d}x^{2}},\\
-k\frac{\text{d}T_{\mathcal{L}}}{\text{d} x}(0,s) = \widetilde{q}_{\mathcal{L}}(s),\quad \frac{\text{d}T_{\mathcal{L}}}{\text{d} x}(L,s) = 0,
\end{gather*}
where $T_{\mathcal{L}}(x,s)$ and $\widetilde{q}_{\mathcal{L}}(s)$ denote the Laplace transforms of $T(x,t)$ and $\widetilde{q}(t)$, respectively. Solving the above problem using standard techniques yields the \revision{following solution in the Laplace domain}:
\begin{gather*}
T_{\mathcal{L}}(x,s) = \frac{T_{0}}{s}+\frac{\widetilde{q}_{\mathcal{L}}(s)[\exp(-mx)+\exp(-m(2L-x))]}{km[1-\exp(-2mL)]},
\end{gather*}
where $m=\sqrt{s(1+\tau s)/\alpha}$. Introducing the geometric series:
\begin{gather*}
\frac{1}{1-\exp(-2mL)} = \sum_{n=0}^{\infty} \exp(-2mnL),
\end{gather*}
and writing $m$ in terms of the difference of two squares
\begin{gather*}
m = \sqrt{\frac{\tau}{\alpha}}\sqrt{\Bigl(s+\frac{1}{2\tau}\Bigr)^{2}-\Bigl(\frac{1}{2\tau}\Bigr)^{2}},
\end{gather*}
yields the following equivalent form of $T_{\mathcal{L}}(x,s)$:
\begin{multline}
 \label{eq:solution_LT} 
T_{\mathcal{L}}(x,s) = \frac{T_{0}}{s}+\frac{\widetilde{q}_{\mathcal{L}}(s)}{k}\sqrt{\frac{\alpha}{\tau}}\left[\Bigl(s+\frac{1}{2\tau}\Bigr)^{2}-\Bigl(\frac{1}{2\tau}\Bigr)^{2}\right]^{-1/2}\left[\sum_{n=0}^{\infty}\exp\left(-\sqrt{\frac{\tau}{\alpha}}(x+2nL)\sqrt{\Bigl(s+\frac{1}{2\tau}\Bigr)^{2}-\Bigl(\frac{1}{2\tau}\Bigr)^{2}}\right)\right.\\ + \left.\sum_{n=0}^{\infty}\exp\left(-\sqrt{\frac{\tau}{\alpha}}(2(n+1)L-x)\sqrt{\Bigl(s+\frac{1}{2\tau}\Bigr)^{2}-\Bigl(\frac{1}{2\tau}\Bigr)^{2}}\right)\right].
\end{multline}
The inverse Laplace transform can now be calculated by applying the convolution theorem and then making use of the following results (see \revision{e.g.} \cite{davison_2004} for the non-standard second result):
\begin{gather*}
\mathcal{L}^{-1}\left\{F(s+a)\right\} = e^{-at}f(t),\\
\mathcal{L}^{-1}\left\{(s^{2}-a^{2})^{-1/2}\exp(-b\sqrt{s^{2}-a^{2}})\right\} = H(t-b)I_{0}(a\sqrt{t^{2}-b^{2}}),
\end{gather*}
where $H(\cdot)$ is the Heaviside function and $I_{0}(\cdot)$ is the modified Bessel function of the first kind of zero order. \revision{Taking the inverse Laplace transform of (\ref{eq:solution_LT}) hence yields:}
\begin{gather*}
\nonumber
\mathcal{L}^{-1}\left\{T_{\mathcal{L}}(x,s)\right\} = T_{0}+\frac{1}{k}\sqrt{\frac{\alpha}{\tau}}\int_{0}^{t}\widetilde{q}(t-u)v(x,u)\,\text{d}u,
\end{gather*}
\revision{where $v(x,u)$ is defined as}
\begin{align}
\nonumber
v(x,u) &= \exp\left(-\frac{u}{2\tau}\right)\left[\sum_{n=0}^{\infty}H\left(u-\sqrt{\frac{\tau}{\alpha}}(x+2nL)\right)I_{0}\left(\frac{1}{2\tau}\sqrt{u^{2}-\frac{\tau}{\alpha}(x+2nL)^{2}}\right)\right.\\ 
\label{eq:v}
&\qquad+ \sum_{n=0}^{\infty}\left.H\left(u-\sqrt{\frac{\tau}{\alpha}}(2(n+1)L-x)\right)I_{0}\left(\frac{1}{2\tau}\sqrt{u^{2}-\frac{\tau}{\alpha}(2(n+1)L-x)^{2}}\right)\right].
\end{align}
Finally, we note that both infinite series in (\ref{eq:v}) can be truncated at a finite number of terms since
\begin{align*}
H\left(u-\sqrt{\frac{\tau}{\alpha}}(x+2nL)\right) &= \begin{cases} 1, & \text{if $n\leq\lfloor (s_{p}t-x)/(2L)\rfloor$},\\ 0, &\text{otherwise},\end{cases}\\
H\left(u-\sqrt{\frac{\tau}{\alpha}}(2(n+1)L-x)\right) &= \begin{cases} 1, & \text{if $n\leq\lfloor (s_{p}t+x)/(2L)-1\rfloor$},\\ 0, &\text{otherwise},\end{cases}
\end{align*}
when $u\in [0,t]$, where $\lfloor\cdot\rfloor$ is the floor function. In summary, we arrive at the following exact solution of the Cattaneo heat-transfer model (\ref{eq:model_pde})--(\ref{eq:model_bcs}):
\begin{gather}
\label{eq:exact_solution}
T(x,t) = T_{0}+\frac{1}{k}\sqrt{\frac{\alpha}{\tau}}\int_{0}^{t}\widetilde{q}(t-u)v(x,u)\,\text{d}u,
\end{gather}
\revision{where $v(x,u)$ is now defined as}
\begin{align}
\nonumber
v(x,u) &= \exp\left(-\frac{u}{2\tau}\right)\left[\sum_{n=0}^{M_{1}}H\left(u-\sqrt{\frac{\tau}{\alpha}}(x+2nL)\right)I_{0}\left(\frac{1}{2\tau}\sqrt{u^{2}-\frac{\tau}{\alpha}(x+2nL)^{2}}\right)\right.\\ 
\label{eq:v_truncated}
&\qquad+ \left.\sum_{n=0}^{M_{2}}H\left(u-\sqrt{\frac{\tau}{\alpha}}(2(n+1)L-x)\right)I_{0}\left(\frac{1}{2\tau}\sqrt{u^{2}-\frac{\tau}{\alpha}(2(n+1)L-x)^{2}}\right)\right],
\end{align}
with $M_{1} = \lfloor (s_{p}t-x)/(2L)\rfloor$ and $M_{2} = \lfloor (s_{p}t+x)/(2L)-1\rfloor$. The above exact solution is useful as it demonstrates that the sample temperature $T(x,t)$ has compact support (up to the additive constant $T_{0}$). In particular, for $t < L\sqrt{\tau/\alpha}$, we see that
\begin{align*}
T(x,t) &= T_{0}+\begin{cases}\displaystyle\frac{1}{k}\sqrt{\frac{\alpha}{\tau}}\int_{0}^{t}\widetilde{q}(t-u)\exp\left(-\frac{u}{2\tau}\right)I_{0}\left(\frac{1}{2\tau}\sqrt{u^{2}-\frac{\tau}{\alpha}x^{2}}\right)\,\text{d}u, & 0 < x < \sqrt{\frac{\alpha}{\tau}}t,\\ 0, & \sqrt{\frac{\alpha}{\tau}}t < x < L,\end{cases}
\end{align*}
which confirms that heat entering the sample at the front-surface is propagated at a finite propagation speed, $s_{p}=\sqrt{\alpha/\tau}$, reaching the back-surface at $t = t_{p} := L\sqrt{\tau/\alpha}$. We remark that the above solution is equivalent to the solution of the Cattaneo heat-transfer model (\ref{eq:model_pde})--(\ref{eq:model_bcs}) on the semi-infinite domain ($L\rightarrow\infty$) with both solutions differing for $t > t_{p}$ when the heat reaches the finite domain boundary at $x=L$ (Figure \ref{fig:1}). Thereafter, for $t > t_{p}$, the exact solution (\ref{eq:exact_solution})--(\ref{eq:v_truncated}) describes right and left travelling thermal waves that move back and forth between the two boundaries at $x=0$ and $x=L$ and decay in amplitude as time progresses \cite{ozisik_1994}.

\subsection{Thermal diffusivity formula}
\label{sec:thermal_diffusivity}
To develop a formula for the thermal diffusivity, we use the rear surface integral method \cite{carr_2019a,carr_2019b,carr_2023a,baba_2009}. This method involves finding a closed-form expression for the rear surface (back surface) integral $\int_{0}^{\infty} T_{\infty}-T(L,t)\,\text{d}t$ (Figure \ref{fig:1})  in terms of the various parameters in the model. This is achieved by considering the generalized function 
\begin{gather}
\label{eq:w1}
w(x) = \int_{0}^{\infty} T_{\infty}-T(x,t)\,\text{d}t.
\end{gather}
In Appendix \ref{app:A}, we show that $w(x)$ satisfies the boundary value problem:
\begin{gather}
\label{eq:w_ode}
\alpha\frac{\text{d}^{2}w}{\text{d}x^{2}} = T_{0}-T_{\infty},\\
\label{eq:w_bcs}
\frac{\text{d}w}{\text{d}x}(0) = \frac{Q_{\infty}}{k},\quad \frac{\text{d}w}{\text{d}x}(L) = 0,\\
\label{eq:w_ac}
\frac{1}{L}\int_{0}^{L}w(x)\,\text{d}x = \int_{0}^{\infty} T_{\infty} - \overline{T}(t)\,\text{d}t,
\end{gather}
while in Appendix \ref{app:B}, we show the following alternative form for $w(x)$ is obtained by solving~(\ref{eq:w_ode})--(\ref{eq:w_ac}):
\begin{gather}
\label{eq:w2}
w(x) = \int_{0}^{\infty} T_{\infty} - \overline{T}(t)\,\text{d}t + \frac{T_{\infty}-T_{0}}{6\alpha}(6Lx-2L^2 - 3x^{2}).
\end{gather}
Equating both (\ref{eq:w1}) and (\ref{eq:w2}) at the back-surface, $x = L$, and rearranging yields the following formula for the thermal diffusivity
\begin{gather*}
\alpha = \frac{L^{2}}{6}\left[\bigintssss_{\,0}^{\infty}\frac{T_{\infty}-T(L,t)}{T_{\infty}-T_{0}}\,\text{d}t - \bigintssss_{\,0}^{\infty} \frac{T_{\infty}-\overline{T}(t)}{T_{\infty}-T_{0}}\,\text{d}t\right]^{-1}.
\end{gather*}
Equivalently, recalling the expressions for $\overline{T}(t)$ (\ref{eq:Tavg}) and $T_{\infty}$ (\ref{eq:Tinf}), we get
\begin{gather}
\label{eq:alpha}
\alpha = \frac{L^{2}}{6}\left[\bigintssss_{\,0}^{\infty}\frac{T_{\infty}-T(L,t)}{T_{\infty}-T_{0}}\,\text{d}t - \bigintssss_{\,0}^{\infty} 1 - \frac{Q(t)}{Q_{\infty}}\,\text{d}t\right]^{-1},
\end{gather}
which is equivalent to the expression obtained for the standard heat equation \cite{carr_2019b}.

\subsection{Relaxation time formula}
\label{sec:relaxation_time}
Recall that the heat pulse applied at the front surface is perceived at the back surface at $t = t_{p} := L\sqrt{\tau/\alpha}$. This result allows the relaxation time to be expressed in terms of the thermal diffusivity:
\begin{gather*}
\tau = \frac{t_{p}^{2}\alpha}{L^{2}},
\end{gather*}
where $t_{p}$ is the time when the back surface temperature first exceeds $T_{0}$, that is, $t_{p} = \min\,\{t > 0\, | \, T(L,t) > T_{0}\}$. Equivalently, using the thermal diffusivity formula (\ref{eq:alpha}), we have
\begin{gather}
\label{eq:tau}
\tau = \frac{t_{p}^{2}}{6}\left[\bigintssss_{\,0}^{\infty}\frac{T_{\infty}-T(L,t)}{T_{\infty}-T_{0}}\,\text{d}t - \bigintssss_{\,0}^{\infty} 1 - \frac{Q(t)}{Q_{\infty}}\,\text{d}t\right]^{-1}.
\end{gather}

\section{Implementation and Verification}
\label{sec:implementation_results}
We now verify the analytical results presented in the previous sections using an illustrative test case. All results are reported using a commonly used parameter set \cite{czel_2013,carr_2019b,carr_2023a}: 
\begin{gather}
\label{eq:parameters1}
k = 222\,\text{W}\,\text{m}^{-1}\text{K}^{-1},\quad\rho = 2700\,\text{kg}\,\text{m}^{-3},\quad c = 896\,\text{J}\,\text{kg}^{-1}\text{K}^{-1},\quad L = 0.002\,\text{m},\\ 
\label{eq:parameters2}
T_{0}=0\,^{\circ}\mathrm{C},\quad q(t) = \frac{Q_{\infty}t}{\beta^{2}}e^{-t/\beta},\quad Q_{\infty} = 7000\,\text{J}\,\text{m}^{-2},\quad \beta = 0.001\,\text{s},
\end{gather}
where $q(t)$ describes an exponential pulse \cite{carr_2019b,carr_2023a} that reaches a peak value at $t = \beta$ (Figure~\ref{fig:1}). The parameters (\ref{eq:parameters1}) yield a target value of the thermal diffusivity (rounded to four decimal places) of
\begin{gather}
\label{eq:alpha_target}
\alpha = \frac{k}{\rho c} = 9.1766\times 10^{-5}\,\text{m}^{2}\text{s}^{-1}.
\end{gather}
For this set of parameter values, the sensitive range for the relaxation time is between $0.0001\,\text{s}$ and $0.001\,\text{s}$ within which we consider four values: 
\begin{gather}
\label{eq:tau_target}
\tau = \text{$0.0001$, $0.004$, $0.007$ or $0.001$}\,\text{s}.
\end{gather}

\begin{figure}[t]
\includegraphics[width=\textwidth]{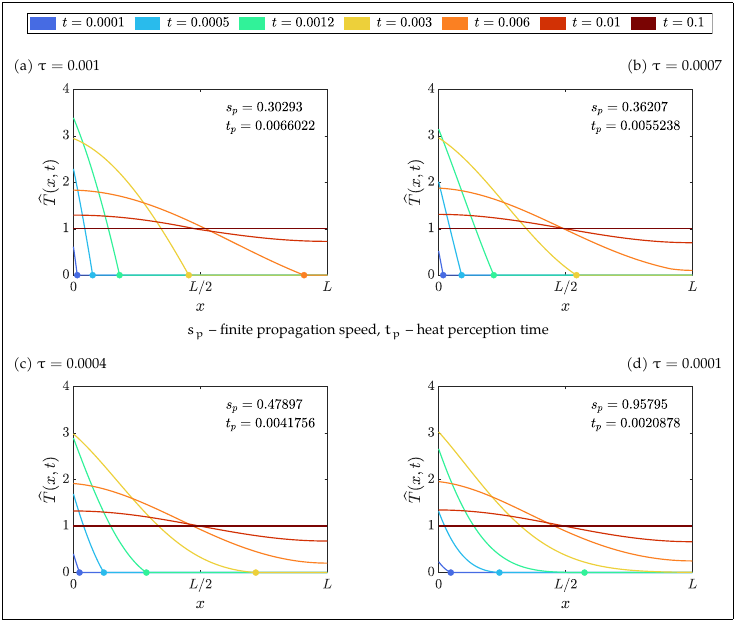}
\caption{\textbf{Effect of relaxation time on temperature field.} Plots of $\widehat{T}(x,t) = (T(x,t)-T_{0})/(T_{\infty}-T_{0})$ [with $T(x,t)$ defined in equations (\ref{eq:exact_solution})--(\ref{eq:v_truncated})] at six different times ($t = 0.0001$, $0.0005$, $0.0012$, $0.003$, $0.006$, $0.01$, $0.1\,\text{s}$) and four different values of the relaxation time $\tau$ [$\mathrm{s}$]: (a) $0.001$ (b) $0.0007$ (c) $0.0004$ (d) $0.0001$. Each figure includes the corresponding values of $s_{p}$ [$\mathrm{m}/\mathrm{s}$] (finite propagation speed) and $t_{p}$ [$\mathrm{s}$] (heat perception time), with the dots denoting the front location at $x = s_{p}t$ when $x \leq L$. All results correspond to the parameter values given in equations (\ref{eq:parameters1})--(\ref{eq:parameters2}) where $L = 0.002\,\mathrm{m}$ and $T_{\infty} = 1.4468\,^{\circ}\mathrm{C}$ (latter rounded to four decimal places). The colour legend at the top of the figure applies to all four panels.}
\label{fig:temperature_fields}
\end{figure}

First we explore the exact solution of the Cattaneo heat-transfer model (\ref{eq:model_pde})--(\ref{eq:model_bcs}). In Figure \ref{fig:temperature_fields}, for each value of $\tau$, we plot $T(x,t)$ (\ref{eq:exact_solution})--(\ref{eq:v_truncated}) and provide numerical values of the finite propagation speed, $s_{p} = \sqrt{\alpha/\tau}$, and heat perception time, $t_{p} = L\sqrt{\tau/\alpha}$. For each value of $\tau$, the front surface temperature increases rapidly before slowing and then decreasing once $q(t)$ exceeds its peak value at $t = \beta = 0.001\,\text{s}$. As time progresses, the heat applied at the front-surface propagates through the sample eventually reaching the back surface at $t = t_{p}$. Thereafter, the back surface temperature increases and the sample approaches thermal equilibrium with the temperature profile at $t = 0.1\,\text{s}$ visually indistinguishable from the steady-state field (\ref{eq:Tinf}). In Figure \ref{fig:temperature_fields}, we see that the relaxation time has a clear observable effect on the temperature profiles with sharper pronounced fronts for the larger values of $\tau$ and smoother less pronounced fronts (resembling those of the standard heat equation) for smaller values of $\tau$. The finite propagation speed and compact support exhibited by solutions to the Cattaneo heat equation is also clearly evident in the temperature profiles. For example, in Figure \ref{fig:temperature_fields}(a), $T(x,t) = 0$ when $x> 0.30293t$ ($x > s_{p}t$) and $t < 0.0066022$ ($t < t_{p}$) with the heat applied at the front surface reaching the back surface at $t = 0.0066022$ (between the $t=0.006$ and $t=0.01$ temperature profiles) while in Figure \ref{fig:temperature_fields}(d), $T(x,t) = 0$ when $x> 0.95795t$ ($x > s_{p}t$) and $t < 0.0020878$ ($t < t_{p}$) with the heat applied at the front surface reaching the back surface at $t = 0.0020878$ (between the $t=0.0012$ and $t=0.003$ temperature profiles).

Next we verify the formulas for the thermal diffusivity and relaxation time derived in the previous section using synthetic discrete temperature data. To generate the synthetic data we (i) evaluate the exact solution $T(x,t)$ (\ref{eq:exact_solution})--(\ref{eq:v_truncated}) at $x = L$ and \revision{at} the parameter values (\ref{eq:parameters1})--(\ref{eq:parameters2}) (ii) sample $T(L,t)$ at intervals of $\Delta t$ (iii) add Gaussian noise to the sampled data. This yields synthetic data taking the form of $N+1$ data points,
\begin{gather}
\label{eq:synthetic_data}
(\widetilde{t}_{0},\widetilde{T}_{0}), (\widetilde{t}_{1},\widetilde{T}_{1}), \hdots, (\widetilde{t}_{N},\widetilde{T}_{N}).
\end{gather} 
Here $\widetilde{t}_{i} = i\Delta t$, $\Delta t = t_{e}/N$, $t_{e}$ is the final sampled time and
\begin{gather*}
\widetilde{T}_{i} = \begin{cases} T_{0}, & \widetilde{t}_{i} < t_{p},\\
T(L,\widetilde{t}_{i}) + \sigma z_{i}, & \widetilde{t}_{i} > t_{p},
\end{cases}
\end{gather*}
where $z_{i}$ is a random number sampled from a normal distribution with mean zero and standard deviation $\sigma$. Using this data, we follow previous work \cite{carr_2019a,carr_2019b,carr_2023a} and approximate the first integral in the thermal diffusivity (\ref{eq:alpha}) and relaxation time (\ref{eq:tau}) formulas using the trapezoidal rule:
\begin{gather}
\label{eq:trap_rule}
\bigintssss_{\,0}^{\infty}\frac{T_{\infty}-T(L,t)}{T_{\infty}-T_{0}}\,\text{d}t \approx \bigintssss_{\,0}^{t_{e}}\frac{T_{\infty}-T(L,t)}{T_{\infty}-T_{0}}\,\text{d}t \approx \Delta t\sum_{i=1}^{N} \left(1 + \frac{2T_{0} - (\widetilde{T}_{i-1}+\widetilde{T}_{i})}{2(T_{\infty}-T_{0})}\right).
\end{gather}
Since the above approximation involves replacing the infinite upper limit with a finite value, thermal equilibrium must be effectively attained at $t = t_{e}$ (i.e. $T(L,t)$ is close to $T_{\infty}$ for $t \geq t_{e}$). For the chosen form of $q(t)$ (\ref{eq:parameters2}), we have an exact expression for the integral involving $Q(t)$ in the thermal diffusivity (\ref{eq:alpha}) and relaxation time (\ref{eq:tau}) formulas:
\begin{gather}
\label{eq:Q_integral}
\bigintssss_{\,0}^{\infty} 1 - \frac{Q(t)}{Q_{\infty}}\,\text{d}t = 2\beta.
\end{gather}
\revision{In summary, substituting (\ref{eq:trap_rule}) and (\ref{eq:Q_integral}) into (\ref{eq:alpha}) and (\ref{eq:tau}) yields} the following estimates for the thermal diffusivity and relaxation time:
\begin{align}
\label{eq:alpha_data}
\widehat{\alpha} &= \frac{L^{2}}{6}\left[\Delta t\sum_{i=1}^{N} \left(1 + \frac{2T_{0} - (\widetilde{T}_{i-1}+\widetilde{T}_{i})}{2(T_{\infty}-T_{0})}\right) - 2\beta\right]^{-1},\\
\label{eq:tau_data}
\widehat{\tau} &= \frac{t_{p}^{2}}{6}\left[\Delta t\sum_{i=1}^{N} \left(1 + \frac{2T_{0} - (\widetilde{T}_{i-1}+\widetilde{T}_{i})}{2(T_{\infty}-T_{0})}\right) - 2\beta\right]^{-1},
\end{align}
where $t_{p} = (t_{m-1}+t_{m})/2$ with $t_{m}$ (\revision{$m\in\{1,\hdots,N\}$}) denoting the minimum discrete time when $\widetilde{T}_{m}$ first deviates from $T_{0}$.

\begin{figure}
\includegraphics[width=\textwidth]{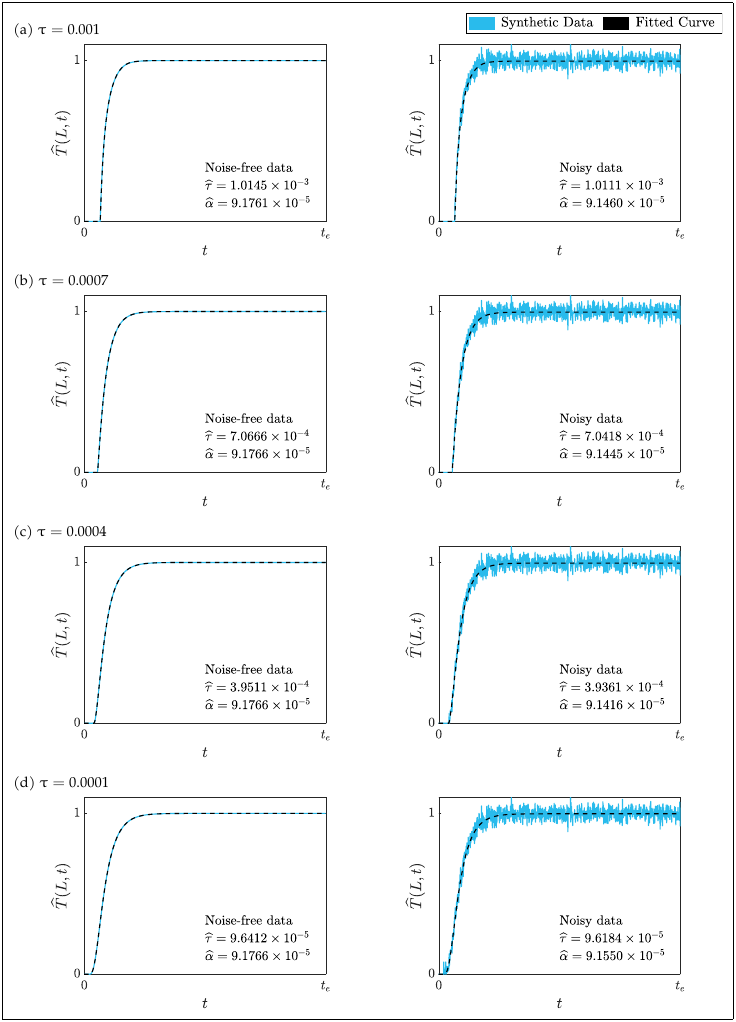}
\caption{\textbf{Verification of thermal diffusivity and relaxation time formulas.} Plots of $\widehat{T}(L,t) = (T(L,t)-T_{0})/(T_{\infty}-T_{0})$,  comparing the synthetic data (solid blue lines) and fitted curve (dashed black lines); with $T(L,t)$ in the latter obtained by evaluating $T(x,t)$ (\ref{eq:exact_solution})--(\ref{eq:v_truncated}) at both $x=L$ and the estimated values of the thermal diffusivity $\widehat{\alpha}$ (\ref{eq:alpha_data}) [$\text{m}^{2}\text{s}^{-1}$] and relaxation time $\widehat{\tau}$ (\ref{eq:tau_data}) [$\text{s}$]. Results are given for noise-free data ($\sigma = 0$, left panels) and noisy data ($\sigma = 0.05\,^{\circ}\mathrm{C}$, right panels) and correspond to the parameter values given in equations (\ref{eq:parameters1})--(\ref{eq:parameters2}) where $t_{e} = 0.1\,\mathrm{s}$ and $T_{\infty} = 1.4468\,^{\circ}\mathrm{C}$ (latter rounded to four decimal places). The colour legend in the top right corner of the figure applies to all eight panels.}
\label{fig:parameterisation_results}
\end{figure}

In Figure \ref{fig:parameterisation_results}, we investigate the accuracy of the thermal diffusivity and relaxation time formulas. Results are reported for both noise-free ($\sigma = 0$) and noisy ($\sigma = 0.05$) synthetic data sets generated using $N = 1000$, $ \Delta t = 10^{-4}\,\text{s}$, $t_{e} = 0.1\,\text{s}$ and one realisation of the random numbers $z_{1},\hdots,z_{N}$ (held constant across all four values of $\tau$). For each value of $\tau$ and each noise level, we plot the synthetic temperature data (\ref{eq:synthetic_data}) and the fitted curve obtained by evaluating $T(x,t)$ (\ref{eq:exact_solution})--(\ref{eq:v_truncated}) at $x = L$ and the computed values of $\alpha$ and $\tau$ (\ref{eq:alpha_data})--(\ref{eq:tau_data}). For the noise-free data, the fitted curve is visually indistinguishable from the synthetic data while for the noisy data, the fitted curve provides an excellent smooth fit of the data. For the noise-free data, we see that the estimated values of the thermal diffusivity agree with the target value (\ref{eq:alpha_target}) to between four and five significant figures. The estimated values of the relaxation time, however, are less accurate due to the accuracy being heavily dependent on the duration of time between temperature samples, $\Delta t$, with the estimated value of $t_{p}$ (c.f. equation (\ref{eq:tau_data})) having error $O(\Delta t)$. As shown in Table \ref{tab:convergence}, decreasing $\Delta t$ improves the accuracy of the relaxation time and demonstrates that it correctly approaches the different specified target values.

\begin{table}[h]
\centering
\begin{tabular}{llllll}
\toprule
& & \multicolumn{2}{l}{$\tau = 0.001$} & \multicolumn{2}{l}{$\tau = 0.0007$}\\
$N$ & $\Delta t$ & $\widehat{\tau}$ & $\widehat{\alpha}$ & $\widehat{\tau}$ & $\widehat{\alpha}$\\
1001 & \num{1.00e-04} & \num{1.0145e-03} & \num{9.1761e-05} & \num{7.0666e-04} & \num{9.1766e-05}\\
10001 & \num{1.00e-05} & \num{1.0008e-03} & \num{9.1766e-05} & \num{7.0030e-04} & \num{9.1766e-05}\\
100001 & \num{1.00e-06} & \num{1.0001e-03} & \num{9.1766e-05} & \num{6.9992e-04} & \num{9.1766e-05}\\
\midrule
& & \multicolumn{2}{l}{$\tau = 0.0004$} & \multicolumn{2}{l}{$\tau = 0.0001$}\\
$N$ & $\Delta t$ & $\widehat{\tau}$ & $\widehat{\alpha}$ & $\widehat{\tau}$ & $\widehat{\alpha}$\\
1001 & \num{1.00e-04} & \num{3.9511e-04} & \num{9.1766e-05} & \num{9.6412e-05} & \num{9.1766e-05}\\
10001 & \num{1.00e-05} & \num{3.9988e-04} & \num{9.1766e-05} & \num{9.9732e-05} & \num{9.1766e-05}\\
100001 & \num{1.00e-06} & \num{3.9998e-04} & \num{9.1766e-05} & \num{9.9971e-05} & \num{9.1766e-05}\\
\bottomrule
\end{tabular}
\caption{\textbf{Convergence of relaxation time estimates.} Estimated values of the thermal diffusivity $\widehat{\alpha}$ (\ref{eq:alpha_data}) [$\text{m}^{2}\text{s}^{-1}$] and relaxation time $\widehat{\tau}$ (\ref{eq:tau_data}) [$\text{s}$] for increasing $N$ (number of samples) and decreasing $\Delta t$ (time between samples) for four target values of the relaxation time ($\tau = 0.0001$, $0.004$, $0.007$, $0.001\,\mathrm{s}$) and one target value of the thermal diffusivity ($\alpha=\num{9.1766e-05}\,\text{m}^{2}\text{s}^{-1}$). All results are for noise-free data ($\sigma = 0$) and the parameter values given in equations (\ref{eq:parameters1})--(\ref{eq:parameters2}).}
\label{tab:convergence}
\end{table}

\begin{figure}[h]
\includegraphics[width=\textwidth]{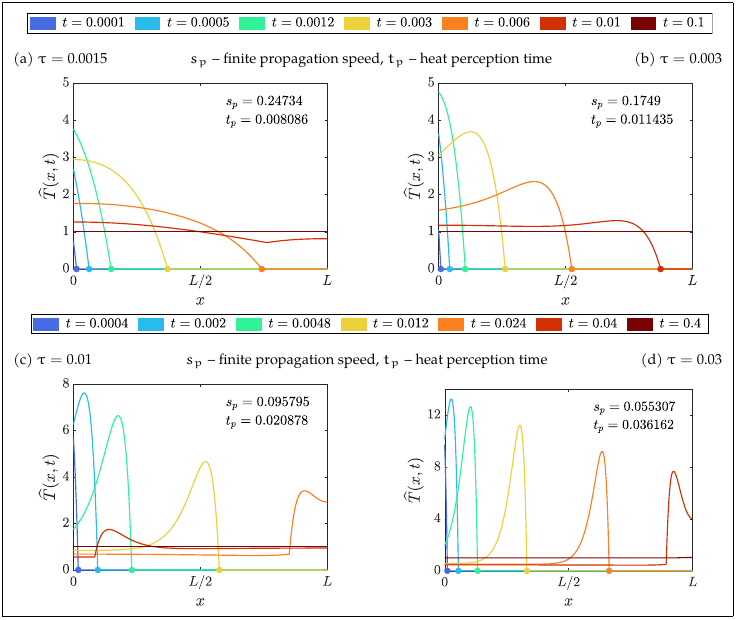}
\caption{\textbf{Effect of larger values of the relaxation time on temperature field.} Plots of $\widehat{T}(x,t) = (T(x,t)-T_{0})/(T_{\infty}-T_{0})$ [with $T(x,t)$ defined in equations (\ref{eq:exact_solution})--(\ref{eq:v_truncated})] at six different times and four different values of the relaxation time $\tau$ [$\mathrm{s}$]: (a) $0.0015$ (b) $0.003$ (c) $0.01$ (d) $0.03$. Each figure includes the corresponding values of $s_{p}$ [$\mathrm{m}/\mathrm{s}$] (finite propagation speed) and $t_{p}$ [$\mathrm{s}$] (heat perception time), with the dots denoting the front location at $x = s_{p}t$ when $x \leq L$. All results correspond to the parameter values given in equations (\ref{eq:parameters1})--(\ref{eq:parameters2}) where $L = 0.002\,\mathrm{m}$ and $T_{\infty} = 1.4468\,^{\circ}\mathrm{C}$ (latter rounded to four decimal places). The top colour legend applies to panels (a) and (b) while the bottom colour legend applies to panels (c) and (d).}
\label{fig:temperature_fields2}
\end{figure}

\begin{figure}
\includegraphics[width=\textwidth]{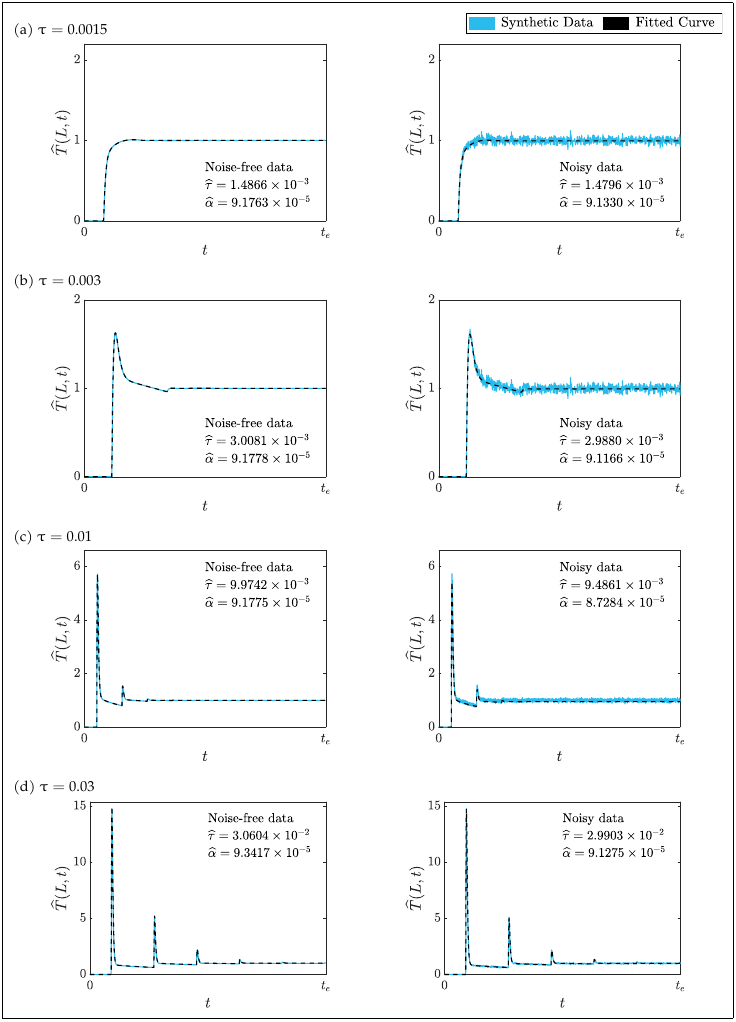}
\caption{\textbf{Verification for larger values of relaxation time.} Plots of $\widehat{T}(L,t) = (T(L,t)-T_{0})/(T_{\infty}-T_{0})$,  comparing the synthetic data (solid blue lines) and fitted curve (dashed black lines); with $T(L,t)$ in the latter obtained by evaluating $T(x,t)$ (\ref{eq:exact_solution})--(\ref{eq:v_truncated}) at both $x=L$ and the estimated values of the thermal diffusivity $\widehat{\alpha}$ (\ref{eq:alpha_data}) [$\text{m}^{2}\text{s}^{-1}$] and relaxation time $\widehat{\tau}$ (\ref{eq:tau_data}) [$\text{s}$]. Results are given for noise-free data ($\sigma = 0$, left panels) and noisy data ($\sigma = 0.05\,^{\circ}\mathrm{C}$, right panels) and correspond to the parameter values given in equations (\ref{eq:parameters1})--(\ref{eq:parameters2}) where $t_{e} = 0.1\,\mathrm{s}$ for $\tau  = 0.0015,0.003$, $t_{e} = 0.4\,\mathrm{s}$ for $\tau  = 0.01,0.03$ and $T_{\infty} = 1.4468\,^{\circ}\mathrm{C}$ (latter rounded to four decimal places). The colour legend in the top right corner of the figure applies to all four panels.}
\label{fig:parameterisation_results2}
\end{figure}

\newpage
\revision{Finally, recall that the exact solution of the Cattaneo heat transfer model (\ref{eq:exact_solution})--(\ref{eq:v_truncated}) exhibits thermal wavefronts propagating back and forth within the sample. For the values of $\tau$ considered in Figure~\ref{fig:temperature_fields}, the amplitudes of these wavefronts are small enough that the back-surface temperature increases monotonically from $T_{0}$ to $T_{\infty}$. However, larger values of $\tau$ yield larger wavefront amplitudes and non-monotone behaviour where the temperature may exceed $T_{\infty}$ or oscillate multiple times before settling to $T_{\infty}$ in the long-time limit. This is well-known behaviour of the Cattaneo equation \cite{ozisik_1994,barletta_1997,kovacs_2015} and while such parameter regimes may be of limited practical use, we note here that the analysis in this work remains valid in such cases as demonstrated in Figures \ref{fig:temperature_fields2} and \ref{fig:parameterisation_results2}}.

\bigskip

\newpage
\revision{\section{Modifications to incorporate heat loss}
\label{sec:heat_loss}
Up until this point we have assumed a perfectly insulated sample (c.f. boundary conditions (\ref{eq:model_bcs})). We now briefly outline an extension of the analysis to incorporate heat losses from the front and back surfaces of the sample (Figure \ref{fig:1}).}

\revision{\subsection{Cattaneo heat transfer model}
To incorporate heat losses, we assume the sample temperature $T(x,t)$ satisfies the heat transfer model (\ref{eq:model_phi_pde})--(\ref{eq:model_phi_bcs}) with modified boundary conditions:
\begin{gather}
\label{eq:model_pde_phi_hl}
\frac{\partial T}{\partial t} + \tau\frac{\partial^{2} T}{\partial t^{2}} = \alpha\frac{\partial^{2} T}{\partial x^{2}},\\
\label{eq:model_ics__phi_hl}
T(x,0) = T_{0},\quad \frac{\partial T}{\partial t}(x,0) = 0,\\ 
\label{eq:model_bcs_phi_hl}
\phi(0,t) = h_{0}(T_{0}-T(0,t)) + q(t),\quad \phi(L,t) = h_{L}(T(L,t)-T_{0}),
\end{gather}
where $h_{0}$ and $h_{L}$ are heat transfer coefficients at the front and back surfaces, respectively. Alternatively, the modified Fourier law (\ref{eq:modified_fourier}) allows us to rewrite the boundary conditions (\ref{eq:model_bcs_phi_hl}), giving the following equivalent model:
\begin{gather}
\label{eq:model_pde_hl}
\frac{\partial T}{\partial t} + \tau\frac{\partial^{2} T}{\partial t^{2}} = \alpha\frac{\partial^{2} T}{\partial x^{2}},\\
\label{eq:model_ics_hl}
T(x,0) = T_{0},\quad \frac{\partial T}{\partial t}(x,0) = 0,\\ 
-k\frac{\partial T}{\partial x}(0,t) = h_{0}(T_{0}-T(0,t)) - \tau h_{0}\frac{\partial T}{\partial t}(0,t) + \widetilde{q}(t),\\ 
\label{eq:model_bcs_hl}
-k\frac{\partial T}{\partial x}(L,t) = h_{L}(T(L,t)-T_{0}) + \tau h_{L}\frac{\partial T}{\partial t}(L,t),
\end{gather}
where $\widetilde{q}(t)$ is as defined earlier in equation (\ref{eq:qtilde}).}

\revision{\subsection{Average temperature}
For the case of heat losses, the average temperature within the sample $\overline{T}(t)$ (\ref{eq:Tavg_def}) does not admit a simple analytical expression nor is it required in any of the analysis that follows, so is not considered further.}

\revision{\subsection{Steady-state temperature}
For the case of heat losses, the steady-state temperature field $\smash{T_{\infty} = \lim\limits_{t\rightarrow\infty} T(x,t)}$ satisfies the boundary value problem:
\begin{gather*}
0 = \alpha\frac{\text{d}^{2} T_{\infty}}{\text{d} x^{2}},\\
-k\frac{\text{d} T_{\infty}}{\text{d} x}(0) = h_{0}(T_{0}-T_{\infty}(0)),\quad -k\frac{\text{d} T_{\infty}}{\text{d} x}(L) = h_{L}(T_{\infty}(L)-T_{0}).
\end{gather*}
which has solution:
\begin{gather}
\label{eq:Tinf_hl}
T_{\infty} = T_{0}. 
\end{gather}
As for the perfectly insulated case (\ref{eq:Tinf}), $T_{\infty}$ is independent of the relaxation time, $\tau$, and is equivalent to the result obtained for the standard heat equation analogue of (\ref{eq:model_pde_hl})--(\ref{eq:model_bcs_hl}).}

\revision{\subsection{Inexact solution for temperature in space and time}
\label{sec:inexact_solution}
To solve the Cattaneo heat transfer model (\ref{eq:model_pde_hl})--(\ref{eq:model_bcs_hl}), we again take Laplace transforms, which yields the following boundary value problem:
\begin{gather*}
(1+\tau s)(sT_{\mathcal{L}}-T_{0}) = \alpha\frac{\text{d}^{2}T_{\mathcal{L}}}{\text{d}x^{2}},\\
-k\frac{\text{d}T_{\mathcal{L}}}{\text{d} x}(0,s) = -h_{0}(1 + \tau s)T_{\mathcal{L}}(0,s) + h_{0}T_{0}(\tau + 1/s) + \widetilde{q}_{\mathcal{L}}(s),\\ 
-k\frac{\text{d}T_{\mathcal{L}}}{\text{d} x}(L,s) = h_{L}(1 + \tau s)T_{\mathcal{L}}(L,s) - h_{L}T_{0}(\tau+1/s).
\end{gather*}
Solving this problem using standard techniques yields the following solution in the Laplace domain:
\begin{gather}
T_{\mathcal{L}}(x,s) = \frac{T_{0}}{s}+\frac{\widetilde{q}_{\mathcal{L}}(s)[(\widetilde{h}_{L}+km)\exp(-mx)-(\widetilde{h}_{L}-km)\exp(-m(2L-x))]}{(\widetilde{h}_{0}+km)(\widetilde{h}_{L}+km) - (\widetilde{h}_{0}-km)(\widetilde{h}_{L}-km)\exp(-2mL)},
\end{gather}
where $m=\sqrt{s(1+\tau s)/\alpha}$, $\widetilde{h}_{0} = h_{0}(1+\tau s)$ and $\widetilde{h}_{L} = h_{L}(1+\tau s)$. Unlike the perfectly insulated case, inverting the Laplace transform back to the time domain is not straightforward analytically. Instead, in the results presented later (section \ref{sec:implementation_results_hl}) we carry out the inversion
\begin{gather}
\label{eq:inexact_solution}
T(x,t) = \mathcal{L}^{-1}\{T_{\mathcal{L}}(x,s)\},
\end{gather}
numerically using the CME (concentrated matrix exponential) method \cite{horvath_2025}.
}

\revision{\subsection{Thermal diffusivity formula}
\label{sec:thermal_diffusivity_hl}
To develop a formula for the thermal diffusivity, we again consider the function (\ref{eq:w1}), which becomes
\begin{gather}
\label{eq:w1_hl}
w(x) = \int_{0}^{\infty} T_{0}-T(x,t)\,\text{d}t,
\end{gather}
since $T_{\infty}=T_{0}$ (\ref{eq:Tinf_hl}). Following similar working to that carried out in Appendix \ref{app:A}, we see that $w(x)$ (\ref{eq:w1_hl}) satisfies the following boundary value problem:
\begin{gather}
\label{eq:w_ode_hl}
\alpha\frac{\text{d}^{2}w}{\text{d}x^{2}} = 0,\\
\label{eq:w_bcs_hl}
-k\frac{\text{d}w}{\text{d}x}(0) + h_{0}w(0) = -Q_{\infty},\quad -k\frac{\text{d}w}{\text{d}x}(L) - h_{L}w(L) = 0,
\end{gather}
which has solution:
\begin{gather}
\label{eq:w2_hl}
w(x) = \frac{h_{L}Q_{\infty}(x-L) - kQ_{\infty}}{k(h_{0}+h_{L}) + h_{0}h_{L}L}.
\end{gather}
Equating both (\ref{eq:w1_hl}) and (\ref{eq:w2_hl}) at the back-surface, $x = L$, and rearranging yields the following modified formula for the thermal diffusivity valid for the case of heat losses:
\begin{gather}
\label{eq:alpha_hl}
\alpha = \frac{h_{0}h_{L}L\int_{0}^{\infty} T(L,t)-T_{0}\,\text{d}t}{\rho c\left[Q_{\infty}-(h_{0}+h_{L})\int_{0}^{\infty} T(L,t)-T_{0}\,\text{d}t\right]}.
\end{gather}
}

\revision{\subsection{Relaxation time formula}
\label{sec:relaxation_time_hl}
For the case of heat losses, heat entering the sample at the front-surface is also propagated at a finite propagation speed, $s_{p} = \sqrt{\alpha/\tau}$, reaching the back-surface at $t = t_{p} := L\sqrt{\tau/\alpha}$. As for the thermal insulated case (section \ref{sec:relaxation_time}), this result allows the relaxation time to be expressed in terms of the thermal diffusivity:
\begin{gather*}
\tau = \frac{t_{p}^{2}\alpha}{L^{2}},
\end{gather*}
where $t_{p}$ is the time when the back surface temperature first exceeds $T_{0}$, that is, $t_{p} = \min\,\{t > 0\, | \, T(L,t) > T_{0}\}$. Equivalently, using the thermal diffusivity formula (\ref{eq:alpha_hl}), we get
\begin{gather}
\label{eq:tau_hl}
\tau = \frac{t_{p}^{2}h_{0}h_{L}\int_{0}^{\infty} T(L,t)-T_{0}\,\text{d}t}{\rho c L\left[Q_{\infty}-(h_{0}+h_{L})\int_{0}^{\infty} T(L,t)-T_{0}\,\text{d}t\right]}.
\end{gather}}

\begin{figure}[h]
\includegraphics[width=\textwidth]{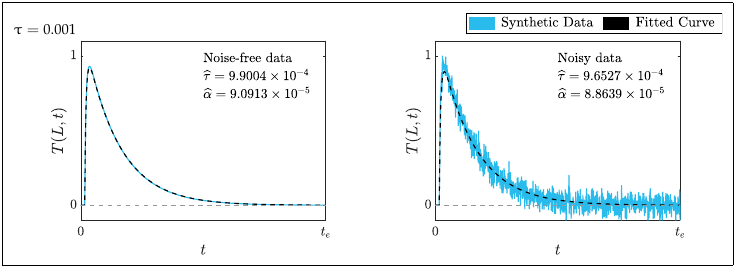}
\caption{\revision{\textbf{Verification of thermal diffusivity and relaxation time formulas (heat losses).} Plots of $T(L,t)$ comparing the synthetic data (solid blue lines) and fitted curve (dashed black lines); with $T(L,t)$ in the latter obtained by evaluating $T(x,t)$ (\ref{eq:inexact_solution}) at both $x=L$ and the estimated values of the thermal diffusivity $\widehat{\alpha}$ (\ref{eq:alpha_hl_data}) [$\text{m}^{2}\text{s}^{-1}$] and relaxation time $\widehat{\tau}$ (\ref{eq:tau_hl_data}) [$\text{s}$]. Results are given for noise-free data ($\sigma = 0$, left panels) and noisy data ($\sigma = 0.05\,^{\circ}\mathrm{C}$, right panels) and correspond to the parameter values (\ref{eq:parameters1})--(\ref{eq:parameters2}) together with $t_{e} = 0.4\,\mathrm{s}$, $h_{0} =  10^{4}\,\text{W}\text{m}^{-2}\text{K}^{-1}$, $h_{L} = 10^{5}\,\text{W}\text{m}^{-2}\text{K}^{-1}$ and $\tau  = 0.001$. The colour legend in the top right corner of the figure applies to both panels.}}
\label{fig:parameterisation_results3}
\end{figure}

\revision{\subsection{Implementation and Verification}
\label{sec:implementation_results_hl}
To apply the thermal diffusivity and relaxation time formulas to the discrete temperature data of the form (\ref{eq:synthetic_data}), we again use the trapezoidal rule to estimate the integral appearing in (\ref{eq:alpha_hl}) and (\ref{eq:tau_hl}):
\begin{gather*}
\int_{0}^{\infty} T(L,t) - T_{0}\,\text{d}t \approx \int_{0}^{t_{e}} T(L,t) - T_{0}\,\text{d}t = \frac{\Delta t}{2}\sum_{i=1}^{N}(\widetilde{T}_{i-1}+\widetilde{T}_{i}-2T_{0}).
\end{gather*}
Inserting this approximation into (\ref{eq:alpha_hl}) and (\ref{eq:tau_hl}) yields the following estimates for the thermal diffusivity and relaxation time:
\begin{align}
\label{eq:alpha_hl_data}
\widehat{\alpha} &= \frac{h_{0}h_{L}L\Delta t\sum_{i=1}^{N}(\widetilde{T}_{i-1}+\widetilde{T}_{i}-2T_{0})}{\rho c\bigl[2Q_{\infty}-(h_{0}+h_{L})\Delta t\sum_{i=1}^{N}(\widetilde{T}_{i-1}+\widetilde{T}_{i}-2T_{0})\bigr]},\\
\label{eq:tau_hl_data}
\widehat{\tau} &= \frac{t_{p}^{2}h_{0}h_{L}\Delta t\sum_{i=1}^{N}(\widetilde{T}_{i-1}+\widetilde{T}_{i}-2T_{0})}{\rho c L\bigl[2Q_{\infty}-(h_{0}+h_{L})\Delta t\sum_{i=1}^{N}(\widetilde{T}_{i-1}+\widetilde{T}_{i}-2T_{0})\bigr]}.
\end{align}
where $t_{p}$ is as defined after equation (\ref{eq:tau_data}). Figure \ref{fig:parameterisation_results3} provides evidence to support the analysis presented in this section for the case of heat losses, with the fitted curves obtained using the above estimates providing excellent fits of the synthetic data.}

\bigskip

\section{Conclusions}
\label{sec:conclusions}
We have studied the laser flash method for measuring thermal diffusivity, carrying out a mathematical analysis to determine the effect of replacing the standard heat equation (which admits \revision{a} non-physical infinite propagation speed) by the Cattaneo heat equation (which exhibits a physical finite propagation speed) in the governing heat transfer model. Key results of the paper include (i) an exact solution of the Cattaneo heat transfer model for the sample temperature across space and time (\ref{eq:exact_solution})--(\ref{eq:v_truncated}), which confirms that heat entering the sample through the front surface reaches the back surface non-instantaneously and (ii) analytical formulas (\ref{eq:alpha})--(\ref{eq:tau}) that express the thermal diffusivity and relaxation time explicitly in terms of the back surface temperature history and the amount of heat absorbed through the front surface. \revision{A somewhat surprising result of the work is that the thermal diffusivity formula remains the same regardless of whether the standard heat equation or the Cattaneo heat equation is used in the analysis. Brief extension of the analysis to incorporate heat losses from the sample was also discussed. 

The work improves the fundamental understanding of the Cattaneo heat equation and provides new formulas that may be useful for model parameterization and material characterization in applications involving non-Fourier heat conduction.} Finally, it is important to note that all analytical results are limited to the specific one-dimensional heat transfer models considered, which assumes that the sample is homogeneous (uniform thermal diffusivity and relaxation time) and subject to uniform heating at the front surface. \revision{One possible direction for future work is to accommodate heterogeneous (layered) samples, which could be achieved} by combining the ideas presented here with those in previous work~\cite{carr_2019b}.

\revision{\section*{Acknowledgements}
The author acknowledges helpful comments received from three anonymous referees, which improved the quality of the final paper.}

\section*{Data Availability}
Supporting MATLAB code \revision{implementing the solutions (sections \ref{sec:exact_solution} and \ref{sec:inexact_solution}), verifying the thermal diffusivity and relaxation time estimates (equations \ref{eq:alpha_data}, \ref{eq:tau_data}, \ref{eq:alpha_hl_data} and \ref{eq:tau_hl_data}), and reproducing the results of the paper} is available on GitHub (\href{https://github.com/elliotcarr/Carr2025a}{https://github.com/elliotcarr/Carr2025a}).

\appendix
\section[Appendix A: Boundary value problem for $w(x)$]{Boundary value problem for $w(x)$}
\label{app:A}
In this appendix, we show that $w(x)$ (\ref{eq:w1}) satisfies the boundary value problem (\ref{eq:w_ode})--(\ref{eq:w_ac}). \revision{First, the} differential equation (\ref{eq:w_ode}) is \revision{obtained} as follows:
\begin{multline*}
\alpha\frac{\text{d}^{2}w}{\text{d}x^{2}} = \alpha\frac{\text{d}^{2}}{\text{d}x^{2}}\int_{0}^{\infty} T_{\infty}-T(x,t)\,\text{d}t = \int_{0}^{\infty} \Bigl(\alpha\frac{\text{d}^{2}T_{\infty}}{\text{d}x^{2}} - \alpha\frac{\partial^{2}T}{\partial x^{2}}\Bigr)\,\text{d}t\\ = - \int_{0}^{\infty}\left(\frac{\partial T}{\partial t}  + \tau\frac{\partial^{2}T}{\partial t^{2}}\right)\,\text{d}t = -\left[T(x,t) + \tau\frac{\partial T}{\partial t}\right]_{0}^{\infty} = T_{0}-T_{\infty},
\end{multline*}
recalling the initial conditions (\ref{eq:model_ics}), steady-state field (\ref{eq:Tinf}) and noting that $\frac{\partial T}{\partial t}(x,t)$ tends to zero in the long time limit. \revision{Next, the} auxiliary condition (\ref{eq:w_ac}) is derived as follows:
\begin{multline*}
\frac{1}{L}\int_{0}^{L} w(x)\,\text{d}x = \frac{1}{L}\int_{0}^{L}\int_{0}^{\infty} T_{\infty}-T(x,t)\,\text{d}t\,\text{d}x = \frac{1}{L}\int_{0}^{\infty}\int_{0}^{L} T_{\infty}-T(x,t)\,\text{d}x\,\text{d}t\\ = \frac{1}{L}\int_{0}^{\infty}T_{\infty}L-\Bigl(\int_{0}^{L}T(x,t)\,\text{d}x\Bigr)\,\text{d}t = \int_{0}^{\infty} T_{\infty} - \overline{T}(t)\,\text{d}t.
\end{multline*}
Finally, the boundary conditions (\ref{eq:w_bcs}) follow directly from \revision{previous work} \cite[Eqs (13) and (14)]{carr_2019b}. 

\section[Appendix B:Alternative solution for $w(x)$]{Alternative solution for $w(x)$}
\label{app:B}
In this appendix, we show \revision{how} the alternative form for $w(x)$ (\ref{eq:w2}) is obtained by solving (\ref{eq:w_ode})--(\ref{eq:w_ac}). \revision{First, we consider the} general solution of the differential equation (\ref{eq:w_ode}) is
\begin{gather}
\label{eq:w_gs}
w(x) = c_{0} + c_{1}x + \frac{(T_{0}-T_{\infty})x^{2}}{2\alpha}.
\end{gather}
\revision{Next, applying} the boundary conditions (\ref{eq:w_bcs}) yields $c_{1} = (T_{\infty}-T_{0})L/\alpha$ when noting that $Q_{\infty} = \rho c L (T_{\infty}-T_{0})$ from (\ref{eq:Tinf}) while applying the auxiliary condition (\ref{eq:w_ac}) yields $c_{0} = \int_{0}^{\infty} T_{\infty} - \overline{T}(t)\,\text{d}t - (T_{\infty}-T_{0})L^{2}/(3\alpha)$. \revision{Finally, substituting} these expressions for $c_{0}$ and $c_{1}$ into (\ref{eq:w_gs}) and simplifying yields the specified form (\ref{eq:w2}).

\footnotesize
\setlength{\bibsep}{1pt plus 0.3ex}

\end{document}